\documentstyle[prd,preprint,aps]{revtex}

\newcommand{\be}{\begin{equation}}
\newcommand{\ee}{\end{equation}}
\newcommand{\bea}{\begin{eqnarray}}
\newcommand{\eea}{\end{eqnarray}}

\begin{document}

\reversemarginpar
\tighten

\title{A follow-up to  `Does 
Nature abhor a logarithm?' \\
(and apparently she doesn't)} 

\author {A.J.M. Medved}

\address{
School of Mathematics, Statistics and Computer Science\\
Victoria University of Wellington\\
PO Box 600, Wellington, New Zealand \\
E-Mail: joey.medved@mcs.vuw.ac.nz}

\maketitle

\begin{abstract}

This letter contains a brief  discussion on 
the leading-order canonical correction to the Bekenstein--Hawking (black hole)
entropy. In particular, we address some 
recent criticism directed at an  earlier commentary.
\end{abstract}

\section*{ }

Let us start by considering
 a rather typical (macroscopically large) Schwarzschild 
black hole in a conventional (four-dimensional) spacetime. 
Let us further stipulate that our black hole has
been emersed in a thermal bath of radiation; with the temperature
being set at precisely the Hawking radiative value \cite{haw}.
Although somewhat contrived,  such a situation
is still quite feasible (in a hypothetical sense)
provided that we can somehow manage to  
enclose the black hole  within
a perfectly reflective ``box'' or, more pragmatically,  agree to adopt
 the model of an anti-de Sitter--Schwarzschild black hole. 
(The latter premise follows
from the anti-de Sitter geometry exhibiting a confining potential that 
effectively
acts like a box; the size of which is determined by the 
anti-de Sitter radius of curvature.)

The utility of the 
above picture  is that it provides us with an idealized  framework
for {\it canonically} modeling the thermodynamic attributes of
a black hole.
Our current interest is specifically with the canonical corrections to the 
entropy of a {\it classically} neutral and static black hole.
(By classically, we actually mean {\it modulo} thermal fluctuations.) 
To further elaborate, the canonical entropy corrections are those which can be
 attributed to  thermal
fluctuations in the black hole horizon area; with a fluctuating area naturally
influencing the entropy via the famed Bekenstein--Hawking relation 
\cite{bek1,haw}. 
(These corrections should not be confused
with the more ``fundamental''   class
of microcanonical corrections
that   arise at the level of state counting. We will return
to this point at  the end of the letter.) 

Let us momentarily suspend belief and suppose  
that, besides the mass, any  other conceivable
``hair'' ({\it i.e.}, macroscopic property) of the black hole has been 
completely and utterly 
``frozen'';  meaning that any
thermally induced fluctuations
in the horizon area can be solely attributed to thermal 
fluctuations in the  mass. In this case, most researchers  will agree that
the leading-order canonical correction to the entropy will go as
$+{1\over 2} \ln S_{BH}$ ({\it e.g.}) \cite{Kastrup,NEW,cmaju}; 
where $S_{BH}$ is the tree-level
or Bekenstein--Hawking value,  which is  (of course) equal to one quarter
of the horizon area in Planck units. 
However, it remains an open question as to
what will be the outcome when the other variables --- most prominently,
the electrostatic charge and the angular
momentum --- are ``turned on'' and also  allowed to fluctuate.
It should be emphasized  that (inasmuch as the horizon area is
functionally dependent  on both the charge and the spin) such 
fluctuations must certainly  
play a role in any realistic type of treatment; irrespective
of whether or not the black hole is, itself, classically
neutral and/or static.

We have found that quantifying the effect of  the angular-momentum 
fluctuations  is 
(technically speaking)  
a particularly
tough nut to crack. So let us again suspend belief and pretend that
the black hole area fluctuates only through its mass
 and (now) its charge. On the  basis of a rigorous grand-canonical treatment,
as documented in \cite{NEW},
this author has recently argued \cite{NEWER} that
a fluctuating charge will induce  an additional  correction
of $+{1\over 2} \ln S_{BH}$ ---  thus leading to a total canonical correction
of $+1 \ln S_{BH}$.  On the other hand, our deduction  has very recently
been criticized  \cite{cmaju2} on the grounds  of a perceived flaw in the
antecedent analysis of \cite{NEW}.  On this point, we will certainly admit
that the {\it presentation} in \cite{NEW}  may have been somewhat misleading;
see the addendum of \cite{NEWER} for a clarification of this issue.
(Note, however, that the criticism, even if  valid, pertains
to only a small part of a rather lengthy paper.) Nonetheless,
 we will now proceed to demonstrate that our conclusion remains intact
by way of a rather simple and intuitive argument.
(For a somewhat related and more formal argument, 
again consult the addendum of 
\cite{NEWER}.) 

So then, how might one go 
about addressing the effect of the charge fluctuations?
Well, as a matter of convenience, we  can just as appropriately ask
what would be the consequence of a fluctuation in the number of 
charged particles
contained inside the black hole. Now consider that, just like a table, a  
six-iron or
a hippopotamus, a black hole (at least one that is formed
out of gravitational collapse) should --- in spite of its own neutrality ---
 consist mainly of 
individually charged  (fermionic) particles. 
Which is to say, if $N$ is the total number of particles
that have been swallowed up by the black hole, then one would naturally
expect the number of charged particles to be given
by (say)  $\eta N$; 
with  $\eta$,  although less than one, definitely being of the order 
of unity. 
There is, in fact, every reason to believe that
$\eta$ would be an essentially universal parameter; that is, 
applicable to most any 
macroscopically
massive object. Hence, it makes just as much sense to work
directly with $N$ and ask what would be the effect of 
fluctuations in the 
particle number.

As it so happens, the impact of a fluctuating particle number
(in addition to a  fluctuating mass)
on the canonically corrected entropy
has already been rigorously addressed in \cite{msett}. The authors of
this paper found a leading-order canonical correction 
(or actually a {\it grand}-canonical
one) of precisely  $+1\ln S_{BH}$; in perfect agreement with
the findings of \cite{NEW} and the claims of \cite{NEWER}. 
As a further point of interest, the analysis of
\cite{msett} was perfectly general in
that it made no presumptions about the
underlying fundamental theory.

But then what about the  fluctuations due to spin?
Well, to reclarify the proposal of \cite{NEWER},
we can justifiably predict a canonical correction of exactly
 $+{1\over 2} \ln S_{BH}$ to be induced by
each macroscopic degree of freedom;
that is, each macroscopic property of the black hole that
can independently influence the horizon area. On the basis of
the above discussion, we might also require   
such a canonical contributor to scale (at least roughly)
with the particle number $N$.
Clearly, most (if not all) particles have an associated  angular momentum,
so that the spin fluctuations should make a similar type of
contribution to that of the charge. Although it is  not 
entirely clear, in the current case,
as to what constitutes an independently fluctuating quantity.
For instance,
 should  each cartesian component of the 
black hole angular momentum be regarded as an independent contributor
or, perhaps, does only  the overall magnitude of
this vector fluctuate independently?  
(This challenging problem will be left for
another day.) 

Glossing over the preceding point, we are still
able to  anticipate  the following form for the canonically corrected entropy:
\be 
S\;=\; S_{BH}\;+\;p\left({1\over 2}\ln[S_{BH}]\right)
\;+\; {\cal O}[S_{BH}^{-1}]\;,
\ee
where  $p$ is meant to represent  
the number of  independently fluctuating  parameters
that  can enter into the black hole area calculation.
Unfortunately, we cannot fix $p$ at this time; indeed, besides
the ambiguity of the spin contribution, there may
be some yet-unknown exotic quantum hair that enters into the fray.
(Even if macroscopically unobservable,  a quantum hair
would still be  relevant as long as it can directly influence  the  
horizon area and can be attributed to  
a significant fraction of existing particles.)
Nevertheless, we can still safely say (by cognizance of the mass, 
charge and spin) that $p\geq3$, so that
the coefficient in front of $\ln S_{BH}$  must be at least three halves.
Moreover, it can reasonably be argued that $p$  should be a {\it universally}
valid parameter once the underlying (quantum-gravitational)  theory 
has been fixed. In this sense, $p$, if it  ever could be measured, may be 
viewed as a direct measure of the quantity of black hole ``hair'' or,
even more esoterically, as  a macroscopic signal from the fundamental
microscopic theory.

As a final comment, let us point out that the total
correction to the black hole entropy should additionally
include the {\it microcanonical} 
contribution \cite{gg2,chatt}, which is usually  attributed to
a  fundamental uncertainty in the 
microstate degeneracy. For instance, some recent
 studies in the context of loop quantum gravity have made a
rather convincing case for a microcanonical correction
of  $ - {1\over 2} \ln S_{BH}$ \cite{mitra,Pol-1,Pol-2}.
Incorporating this result into the current (canonical) discussion,
we can now  expect a {\it total} logarithmic correction of at
least  $+1\ln S_{BH}$.
On the basis of such a lower bound, this author must concede that 
it is  quite unlikely that the canonical and microcanonical corrections 
 could generically  cancel; a possibility that was speculated upon 
in our earlier article 
\cite{NEWER}.  
Which is to say, in answer to the title of \cite{NEWER},
 Nature most probably does {\it not} 
abhor logarithmic corrections (but may
yet have a grievance with the people who write papers about them.)

\section*{Acknowledgments} 

 Research is supported  by
the Marsden Fund (c/o the  New Zealand Royal Society) 
and by the University Research  Fund (c/o Victoria University).
The author expresses his gratitude to Magrietha Medved for
proof reading and moral support.

\end{document}